\documentclass[12pt]{article}

\usepackage{graphicx}
\usepackage{dcolumn}
\usepackage{bm}
\usepackage{color}

\begin{document}

\title{Turbulence generation by a shock wave interacting with a random density inhomogeneity field}

\author{C. Huete Ruiz de Lira}

\maketitle

\begin{abstract}
When a planar shock wave interacts with a random pattern of pre-shock density non-uniformities, it generates an anisotropic turbulent velocity/vorticity field. This turbulence plays an important role at the early stages of the mixing process in the compressed fluid. This situation emerges naturally in shock interaction with weakly inhomogeneous deuterium-wicked foam targets in Inertial Confinement Fusion (ICF) and with density clumps/clouds in astrophysics. We present an exact small-amplitude linear theory describing such interaction. It is based on the exact theory of time and space evolution of the perturbed quantities behind a corrugated shock front for a single-mode pre-shock non-uniformity. Appropriate mode averaging in 2D results in closed analytical expressions for the turbulent kinetic energy, degree of anisotropy of velocity and vorticity fields in the shocked fluid, shock amplification of the density non-uniformity, and sonic energy flux radiated downstream. These explicit formulas are further simplified in the important asymptotic limits of weak/strong shocks and highly compressible fluids. A comparison with the related problem of a shock interacting with a pre-shock isotropic vorticity field is also presented.
\end{abstract}

\section{Introduction}
The interaction of shock waves with flow inhomogeneities attracted the attention of scientists working in different fields, ranging from Aerodynamics \cite{5,6,7,8,9,10,18,19}, Shock tube research \cite{1,2,3}, Laser Fusion \cite{12,16,17,22,25,26,27,28,29,30,31,32,33,34,35} and Astrophysics \cite{36,37,37b}. The physics of the interaction has been described theoretically \cite{5,6,7,8,9,10,11,12} and experimentally \cite{23,24,24b}. In the last decades, the advent of super computers provided a sophisticated and accurate tool with which to follow the details of that interaction \cite{12,13,14,15,16,20,21,22}. However, the development of analytical works is an important and complementary way to guide in the design of numerical simulations and experiments. This work is a natural continuation of a previous article \cite{17}, in which the interaction of a planar shock with a pre-shock turbulent vorticity perturbation field was considered, with the help of an analytical theory.
The shock/density field interaction has been studied in the recent past by different researches with the help of numerical simulations. Rotman \cite{12} provided results of that study assuming 2D perturbations upstream of the shock and Mahesh \textit{et al} \cite{13} studied a full 3D spectrum of steady state density perturbations. With the help of 2D Large Eddy Simulations (LES), Rotman described the interaction of a shock wave with random vorticity and density fields separately. For both cases, he noticed that shock compression waves greatly amplifies the upstream turbulence and reduces the turbulent length scales \cite{12}. On the other hand, Mahesh \textit{et al} have used Direct Numerical Simulations (DNS) to study the interaction of a combined spectrum of pre-shock vorticity and density inhomogeneities. Besides, the problem considered here is of importance in the field of Inertial Confinement Fusion (ICF), where a sequence of well tuned shocks must be launched inside the thermonuclear foam targets. The use of foams has been proposed over 20 years ago to improve the stability of the implosion and give an effective smoothing of the laser energy deposition process \cite{28,29,30,31,32,33,34,35}. However, given the non uniform character of the background density through which the first shock travels inside the target, it is clear that the shock front will dynamically react against those perturbations, generating additional density, vorticity and acoustic fluctuations in the compressed fluid, much in the same way as happens in the shock/vorticity interaction studied in \cite{17} and references therein.
The anisotropic turbulent spectrum left by the shock front can act as a trigger for further hydrodynamic instabilities and mixing.
\begin{figure} [ht]
\centering
\begin{center}
\vspace*{-3mm}
\includegraphics[width= 0.5 \textwidth, clip]{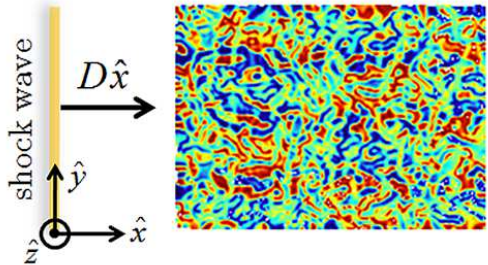}
\end{center}
\vspace*{- 6 mm}
\caption{A planar shock moving with velocity $D\hat{x}$ in the laboratory reference frame hits a 2D turbulent density field located in the half-space $x'\geq0$.}
\label{Figure-1}
\end{figure}
In this work we present a completely analytical model to study the outcome of the interaction of a planar shock front with a non uniform pre-shock 2D density profile, continuing the task started in \cite{17}. The strategy is simple: an arbitrary pre-shock spectrum may be decomposed in its Fourier modes and the shock interaction with every single mode is studied separately and superposition is later performed to see the effect on the whole spectrum. Each shock/single mode interaction is assumed to be linear, that is, the initial density fluctuation is assumed to be much lower than the mean background density, a condition satisfied in ICF targets. Statistical averages are obtained by integrating over the angles that define the orientation of the pre-shock perturbation wavenumber vector in space, similarly as has been done in Refs.\cite{14,17} and the works cited there. We will restrict our results to isotropic pre-shock spectra.
This work is structured as follows: in the following Section 2, the 2D single mode theory is developed, and we show the exact shock dynamics with its corresponding asymptotic expressions. In Section 3 the 2D isotropic random density field is considered. We obtain exact expressions for the kinetic energy and vorticity generation and density amplification, and acoustic energy flux emitted by the shock.
\section{Interaction of a planar shock with a single mode 2D density field}
\subsection{Wave Equation and Boundary Conditions}
\begin{figure} [ht]
\centering
\begin{center}
\vspace*{-3mm}
\includegraphics[width= 0.9 \textwidth, clip]{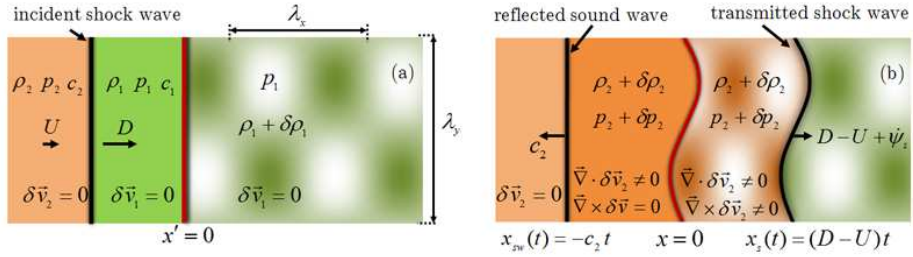}
\end{center}
\vspace*{- 6 mm}
\caption{(a) A planar shock wave travels with velocity $D$ at the laboratory reference frame before enters into the perturbed density field . (b) The transmitted corrugated shock moves into the disturbed fluid at the compressed fluid frame for $t>0$ with velocity $(D-U)$. Density, vorticity and acoustic fluctuations are generated behind it. A stable sound wave is reflected to the left traveling with velocity $c_2$ in this frame.}
\label{Figure-2}
\end{figure}

A planar shock is incident at $t=0$ at the surface $x'=0$ in the laboratory frame. The fluid is an ideal gas with adiabatic exponent $\gamma$ and the perturbations consist of a weakly inhomogeneous density field. The shock comes from the left $(x'=-\infty)$ and travels with velocity $D\hat{x}'$, as measured in the laboratory reference system. In the uniform half-space $x'<0$. the density and pressure ahead the shock are $\rho_1$, $p_1$ respectively, and $\rho_2$, $p_2$ are the values behind it. The velocity of the compressed fluid is $U\hat{x}'$, also measured in the laboratory frame reference. The upstream sound speed is $c_1$, and the downstream value is $c_2$. The shock Mach number with respect to the upstream gas is $M_1=D/c_1\geq1$  and the shock Mach number with respect to the compressed fluid is $M_2=(D-U)/c_2\leq1$. Before the shock arrives to the interface $x'=0$, the relationship between the quantities at both sides the shock front are:
\begin{equation}
R=\frac{\rho_2}{\rho_2}=\frac{D}{D-U}=\frac{(\gamma+1)M_1^2}{(\gamma-1)M_1^2+2}
\ , \label{1}
\end{equation}
\begin{equation}
M_2=\frac{D-U}{c_2}=\sqrt{\frac{(\gamma-1)M_1^2+2}{2\gamma M_1^2-\gamma+1}}
\ , \label{2}
\end{equation}
\begin{equation}
\frac{p_2}{p_1}=\frac{2\gamma M_1^2-\gamma+1}{\gamma+1}
\ , \label{3}
\end{equation}
\begin{equation}
\frac{c_2}{c_1}=\frac{\sqrt{(2\gamma M_1^2-\gamma+1)[(\gamma-1)M_1^2+2]}}{(\gamma+1)M_1}
\ . \label{4}
\end{equation}
In the right half-space, the perturbed density field is described by $\delta \rho_1(x',y)= \rho_1\epsilon_k\cos(k_x x')\cos(k_y y)$ where $x'$ and $y$ are the longitudinal and transverse coordinates as measured in the laboratory system of reference. To remain within the limits of validity of the linear theory, we assume $\epsilon_k\ll1$, where $\epsilon_k $ is a function of $k=\sqrt{k_x^2+k_y^2}$, as isotropy is assumed for the pre-shock perturbations. The longitudinal and transverse wave numbers are defined, respectively by: $k_x=2\pi/\lambda_x$ and $k_y=2\pi/\lambda_y$, with $\lambda_x$ and $\lambda_y$ characteristic lengths.
Once the shock is in the half-space $x'\geq0$, the density profile in front of it will induce density, pressure and velocity fluctuations downstream, and its shape will be distorted. We assume that all the perturbed quantities are much smaller than the background values. We denote the upstream values with the subscript 1 and the downstream values with the subscript 2. From now on, the linearized equations of motion will be written and solved in a system of reference that co-moves with the compressed fluid particles. We define the followings dimensionless perturbation functions, factoring out the small parameter $\epsilon_k$:
\begin{eqnarray}
\frac{\delta\rho_2}{\rho_2}=\epsilon_k\tilde{\rho}(x,t)\cos(k_y y)\ , \nonumber \\
\frac{\delta p_2}{\rho_2 c_2^2}=\epsilon_k\tilde{p}(x,t)\cos(k_y y)\ , \nonumber \\
\frac{\delta v_{2x}}{c_2}=\epsilon_k\tilde{v}_x(x,t)\cos(k_y y)\ , \nonumber \\
\frac{\delta v_{2y}}{c_2}=\epsilon_k\tilde{v}_y(x,t)\sin(k_y y)
\ . \label{5}
\end{eqnarray}
In Eq.(\ref{5}) $t$ is the time, and $x$ is the longitudinal coordinate as measured in the compressed fluid frame. The quantities $\tilde{v}_x$ and $\tilde{v}_y$ correspond to the longitudinal and transverse velocities respectively, and $\tilde{\rho}$ and $\tilde{p}$ represent the dimensionless density and pressure perturbations. We also define the dimensionless time $\tau = k_y c_2 t$.
The linearized equations of motion in the compressed fluid frame are:
\begin{eqnarray}
\frac{\partial \tilde{\rho}}{\partial \tau} &=& -\frac{\partial \tilde{v}_x}{\partial \left( k_y \ x\right)} -  \tilde{v}_y \ , \nonumber \\
\frac{\partial \tilde{v}_x}{\partial \tau} &=& -\frac{\partial \tilde{p}}{\partial \left(k_y \ x\right)} \ , \nonumber \\
\frac{\partial \tilde{v}_y}{\partial \tau} &=&  \tilde{p} \ ,
\label{6}
\end{eqnarray}
representing the mass, \textit{x}-momentum and \textit{y}-momentum conservation equations, respectively. Furthermore, the conservation of entropy holds if we assume adiabatic flow behind the shock, which is represented by:
\begin{equation}
\frac{\partial \tilde{p}}{\partial \tau} = \frac{\partial \tilde{\rho}}{\partial \tau}.
\label{7}
\end{equation}
The dynamics of the perturbed quantities in the whole compressed fluid is governed by the wave equation. Combining Eqs.(\ref{6}) and (\ref{7}) we get the wave equation for the pressure fluctuations:
\begin{equation}
\frac{\partial^2 \tilde{p}}{\partial \tau^2} = \frac{\partial^2 \tilde{p}}{\partial (k_y x)^2}-\tilde{p}.
\label{8}
\end{equation}
We assume that the shock front hits the interface $x=0$ at $t=0$. At $t=0^+$, a shock is transmitted to the right into the perturbed half-space and a neutrally stable sound wave is reflected back inside the region $x<0$, as shown in Fig.\ref{Figure-2}. As the shock wave travels in a non-uniform fluid, the shock will be distorted. We define the shock ripple $\psi_s(y,t)$ as the deviation from planarity. The shock ripple oscillates in time, generating pressure fluctuations that propagate with the local sound speed into the compressed fluid. The sound waves generated by the shock oscillation can be stable or evanescent waves, depending on the ratio $k_x/k_y$, the shock Mach number $M_1$ and the gas compressibility $\gamma$ \cite{17}. We assume that no sound wave hits the shock surface from behind (isolated shock). At the surface $x=0$, pressure and normal velocity are continuous on both sides of it. However, the distorted front generates vorticity and entropy perturbations and the neutral sound wave to the left does not. Hence, the \textit{x}-derivative of $\tilde{v}_y$ and density are generally discontinuous at $x=0$. It is not difficult to get the following relationships just to the right of the left traveling sound wave:
\begin{eqnarray}
\tilde{v}_x+\tilde{p} =0 \ , \nonumber \\
\tilde{v}_y=0  \ .
\label{9}
\end{eqnarray}
The boundary conditions at the shock are obtained after linearizing Rankine-Hugoniot conditions and using the continuity of the tangential velocity. We write them here for the particular case of density/entropy pre-shock modulation at constant pressure and zero velocity perturbation ahead of the shock:
\begin{equation}
\frac{d \xi_s}{d\tau} =\frac{\gamma+1}{4 M_2} \tilde{p} - \frac{M_2 R}{2} \frac{\delta\rho_1}{\rho_1},
\label{10}
\end{equation}
\begin{equation}
\tilde{v}_x = \frac{M_1^2+1}{2 M_1^2 M_2}\tilde{p} - \frac{M_2(R-1)}{2}\frac{\delta\rho_1}{\rho_1},
\label{11}
\end{equation}
\begin{equation}
\tilde{v}_y = M_2 (R-1) \xi_s\ .
\label{12}
\end{equation}
\begin{equation}
\tilde{\rho} = \frac{1}{M_1^2 M_2^2}\tilde{p}+\frac{\delta\rho_1}{\rho_1},
\label{13}
\end{equation}
where Eq.(\ref{10}) represents the mass equation, Eqs.(\ref{11}) and (\ref{12}) correspond to the longitudinal and transverse momentum conservation respectively, and Eq.(\ref{13}) is the energy equation. Here, $\xi_s \epsilon_k= k_y \psi_s$ is the dimensionless shock ripple amplitude. To get the perturbed quantities in the compressed fluid we solve the wave equation [Eq.(\ref{8})] with adequate boundary conditions. To this scope, we concentrate at the weak discontinuity $(x=0)$ and at the shock front $[x=x_s(t)=(D-U)t]$. To finally solve the dynamics of the shock front, we need the initial value of the pressure perturbation behind the shock. Using Eqs.(\ref{9}), (\ref{11}) and (\ref{13}) it is easy to get \cite{16}:
\begin{equation}
\tilde{p}_{s0} = \frac{M_1^2 M_2^2 (R-1)}{2 M_1^2 M_2 + M_1^2 + 1},
\label{14}
\end{equation}
where $\tilde{p}_{s0}$ is the initial shock pressure perturbation at $t=0^+$. Besides, it is clear that the initial shock ripple amplitude is $\xi_{s0}\equiv0$, as the shock front is planar in shape when it arrives to $x=0$.
\subsection{Pressure dynamics in the compressed fluid}
To solve the wave equation [Eq.(\ref{8})] inside the compressed fluid, we use the following coordinate transformation \cite{4,16,17,26,27}:
\begin{eqnarray}
\nonumber
k_y x = r \sinh \chi,  \\
\tau = r \cosh \chi.
\label{15}
\end{eqnarray}
Here, $\chi=const$ represents a planar front defined by $x=c_2 t \tanh{\chi}$. The shock front coordinate is given by: $\tanh{\chi_s}=M_2$, and from Eq.(\ref{15}) we get:
\begin{equation}
\tau=r_s\cosh{\chi_s}=\frac{r_s}{\sqrt{1-M_2^2}}
\label{16}
\end{equation}
The wave equation [Eq.(\ref{8})] is now rewritten as:
\begin{equation}
r^2\frac{\partial^2 \tilde{p}}{\partial r^2} +r \frac{\partial \tilde{p}}{\partial r} + r^2\tilde{p} = \frac{\partial^2 \tilde{p}}{\partial \chi^2}.
\label{17}
\end{equation}
The boundary conditions at the shock front [Eqs.(\ref{10})-(\ref{13})] can be recast as:
\begin{equation}
\frac{1}{r}\frac{\partial \tilde{p}}{\partial \chi}\left.{\!\!\frac{}{}}\right)_{\chi_s} = -\frac{M_1^2 + 1}{2 M_1^2 M_2}\frac{d \tilde{p}_s}{d r_s} - \frac{M_2^2 (R-1)}{\sqrt{1-M_2^2}} \xi_s - \frac{M_2(R-1)}{2}\zeta_0 \sin{(\zeta_0 r_s)}
\ , \label{18}
\end{equation}
\begin{equation}
\frac{d\xi_s}{dr_s} = \frac{\gamma + 1}{4 M_2\sqrt{1-M_2^2} } \tilde{p_s} -\frac{M_2 R}{2\sqrt{1-M_2^2}} \cos{(\zeta_0 r_s)}
\ ,  \label{19}
\end{equation}
where $\zeta_0$ is a dimensionless frequency that characterizes the periodicity of the pre-shock density inhomogeneity. Its value is given by:
\begin{equation}
\zeta_0=\frac{k_x}{k_y}\frac{M_2 R}{\sqrt{1-M_2^2}}
\ . \label{20}
\end{equation}
We solve Eqs.(\ref{17})-(\ref{19}) by using the Laplace transform. For any quantity $\varphi(\chi,r)$ we define its Laplace transform by: $\Phi(\chi,s)=\int_0^\infty\varphi(\chi,r)\exp(-sr)dr$. After some algebra which we omit here, but that can be found in \cite{16,17}, we get an exact closed form expression for the Laplace transform of the shock front pressure fluctuations $\tilde{P}_s$:
\begin{eqnarray}
&\tilde{P}_s(s)& = \frac{M_1^2 M_2^2 (R-1) s}{2 M_1^2 M_2s\sqrt{s^2+1} + \left( M_1^2+1 \right)s^2+M_1^2}+ \nonumber \\
&+& \frac{ 2M_1^2 M_2 \alpha_e  s}{\left[ 2 M_1^2 M_2s\sqrt{s^2+1} + \left( M_1^2+1 \right)s^2+M_1^2 \right]\left( s^2+\zeta_0^2 \right)}
\ .  \label{21}
\end{eqnarray}
The coefficient $\alpha_e$, which is the same as $\alpha_{20}$ in \cite{16} is given by:
\begin{equation}
\alpha_e = \frac{M_2 (R-1)}{2}\left(\frac{M_2^2 R}{1-M_2^2}-\zeta_0^2 \right)
\ , \label{22}
\end{equation}
The evolution of the shock pressure perturbation as a function of the time is obtained by calculating the inverse Laplace transform of Eq.(\ref{21}), after an integration in the complex plane. We formally write \cite{38,39,40}:
\begin{equation}
\tilde{p}_s(r_s)=\frac{1}{2\pi i}\int_{c-i \infty}^{c+i \infty}\tilde{P}_s(s)\exp(s r_s)ds
\ , \label{23}
\end{equation}
It is clear that the denominator of Eq.(\ref{21}) contributes with stable asymptotic oscillations of frequency $\zeta_0$. After analyzing the residues at the poles $s=\pm i \zeta_0$, we get the asymptotic expressions, similarity as in Ref.\cite{17}:
\begin{equation}
\tilde{p}_s(\tau\gg1) \cong  \left \{ \begin{array}{ll}
	e_{lr} \cos \left(\ \zeta_0 r_s\ \right) + e_{li} \sin \left(\ \zeta_0 r_s\ \right) ,\ \ \zeta_0 \leq 1 \\
				 \\
	e_s \cos \left(\ \zeta_0 r_s\ \right), \ \ \ \ \ \ \ \ \ \ \ \ \ \ \  \ \ \ \ \ \ ,\ \ \zeta_0 \geq 1  \\
	\end{array} \right.  \ \ ,
\label{24}
\end{equation}
where $r_s=\sqrt{1-M_2^2}\ \tau$, and the coefficients $e_{lr}$, $e_{li}$ and $e_{s}$ are the same as those obtained in \cite{16}, and formally equivalent to those shown in Eq.(47) in \cite{17}.
\begin{eqnarray}
e_{lr}=\frac{2M_1^2 M_2 \left[  M_1^2 -\left(  M_1^2+1   \right) \zeta_0^2  \right]\alpha_e }{4M_1^4 M_2^2 \zeta_0^2 \left( 1-\zeta_0^2  \right)+ \left[  M_1^2 -\left(  M_1^2+1   \right) \zeta_0^2  \right]^2}
\ , \nonumber \\
e_{li} = \frac{4 M_1^4 M_2^2 \zeta_0 \sqrt{1-\zeta_0^2}\  \alpha_e }{4M_1^4 M_2^2 \zeta_0^2 \left( 1-\zeta_0^2  \right)+\left[  M_1^2 -\left(  M_1^2+1   \right) \zeta_0^2  \right]^2}
\ , \nonumber \\
e_s = -\frac{2M_1^2M_2 \ \alpha_e }{2 M_1^2 M_2 \zeta_0\sqrt{\zeta_0^2 - 1} + \left( M_1^2 + 1  \right)\zeta_0^2 - M_1^2}
\ . \label{25}
\end{eqnarray}
The shock front ripple oscillates with the dimensionless frequency $\zeta_0$ within the domain of the variable $r_s$. If $\zeta_0>1$ sound waves fill the space behind it in the form of traveling fronts. At any position $x$, the sound waves will reach the asymptotic regime when the shock is far enough $(x_s\gg x)$. Because of the Doppler effect, the dimensionless frequency of the compressed fluid particles oscillations is $\zeta_1<\zeta_0$. The value of $\zeta_1$ can be seen to be given by \cite{16,17}:
\begin{equation}
\zeta_1=\frac{\zeta_0-M_2\sqrt{\zeta_0^2-1}}{\sqrt{1-M_2^2}}
\ . \label{26}
\end{equation}
Besides, a longitudinal wave number $k_x^{ac}$ is associated to the stable fluid oscillations given by \cite{16,17}:
\begin{equation}
\frac{k_x^{ac}}{k_y}=\frac{M_2 \zeta_0-\sqrt{\zeta_0^2-1}}{\sqrt{1-M_2^2}}
\ . \label{27}
\end{equation}
It is not difficult to see that the asymptotic pressure oscillation at any position $x$, is therefore given by:
\begin{equation}
\tilde{p}(x,\tau)=e_s\ \cos{(\zeta_1 \tau -k_x^{ac}x)}
\ . \label{28}
\end{equation}
For $\zeta_0<1$, the sound waves emitted by the shock front are evanescent and decay exponentially away from it \cite{16,17}. From Eq.(\ref{27}) it is easy to see that for $1\leq\zeta_0\leq 1/\sqrt{1-M_2^2}$, the waves are emitted to the right, following the shock front, and for $\zeta_0> 1/\sqrt{1-M_2^2}$ the sound waves escape to the left, filling the whole compressed fluid.
\subsection{Rotational and irrotational perturbations downstream}
As the shock ripple oscillates in time, not only pressure oscillations are generated downstream, but also vorticity and velocity fluctuations. The vorticity is strictly generated at the shock front and remains frozen to the fluid elements in the absence of viscosity. Let us define the 2D dimensionless gradient operator:
\begin{equation}
\vec{\nabla}_{2D}=\left(\frac{\partial}{\partial k_y x}\ ,\ \frac{\partial}{\partial k_y y} \right)
\ . \label{29}
\end{equation}
It is clear that the vorticity is directed along the z-axis and is defined by:
\begin{equation}
\vec{\omega}=k_y c_2 \left(\vec{\nabla}_{2D}\ \times\ \tilde{\vec{v}}\right)
\ , \label{30}
\end{equation}
with $\tilde{\vec{v}}=(\tilde{v}_x,\tilde{v}_y)$.
The function $\vec{\omega}$ can be found by using the continuity of tangential velocity at the shock front and that vorticity is conserved along the particles paths. Following \cite{17,26} we get:
\begin{equation}
\tilde{\vec{\omega}}(x,y)=\left[\Omega_2 \tilde{p}_s\left(r_s=\frac{k_y x \sqrt{1-M_2^2}}{M_2}\right)+\Omega_3\cos{(Rk_xx)}\right]\sin{(k_yy)}
\ , \label{31}
\end{equation}
where $\Omega_2$ and $\Omega_3$ are given by:
\begin{equation}
\Omega_2 = \frac{\left( M_1^2-1  \right)^2 \sqrt{2 \gamma M_1^2 - \gamma +1}}{M_1^2 \left[  \left( \gamma -1 \right) M_1^2 + 2 \right]^{3/2}}
\ , \label{32}
\end{equation}
\begin{equation}
\Omega_3 = -\frac{M_2(R^2-1)}{2}
\ . \label{33}
\end{equation}
The first term, proportional to $\Omega_2$ is generated by the shock front distortion (or shock curvature as named in \cite{18,19}). It is always present whenever the shock front gets corrugated as in any RMI like problem \cite{17}. The second term originated from the interaction between the pre-shock density field and the zero order pressure jump across the shock surface, usually called the baroclinic term \cite{18,19}. We omit a term $\Omega_1$ in Eq.(\ref{31}) that would appear for a pure pre-shock vorticity field ahead of the shock wave, a case that has been studied in detail in \cite{17}.
After combining Eqs.(\ref{6}) and (\ref{7}), the velocity field downstream can be seen to satisfy the differential equation \cite{17}:
\begin{equation}
\frac{\partial^2 \tilde{\vec{v}}}{\partial \tau^2} = \vec{\nabla}_{2D}^2\tilde{\vec{v}} + \vec{\nabla}_{2D} \times\left(\vec{\nabla}_{2D} \times \tilde{\vec{v}}\right)
\ . \label{34}
\end{equation}
To solve it, we decompose the velocity into a rotational and acoustic component:
\begin{equation}
\tilde{\vec{v}}(x,y,t) = \tilde{\vec{v}}_{rot}(x,y)+\tilde{\vec{v}}_{ac}(x,y,t)
\ . \label{35}
\end{equation}
The rotational contribution, which accounts for the vorticity downstream, is time independent in the compressed fluid frame, and it satisfies:
\begin{equation}
\vec{\nabla}_{2D}^2\tilde{\vec{v}}_{rot}=  - \vec{\nabla}_{2D} \times\left(\vec{\nabla}_{2D} \times \tilde{\vec{v}}_{rot}\right)
\ , \label{36}
\end{equation}
and the acoustic part satisfies the homogeneous wave equation:
\begin{equation}
\frac{\partial^2 \tilde{\vec{v}}_{ac}}{\partial \tau^2} = \vec{\nabla}_{2D}^2\tilde{\vec{v}}_{ac}
\ . \label{37}
\end{equation}
An exact solution can be sought for both contributions in the same way as has been done in \cite{17} for the pre-shock vorticity case. As we are interested in the asymptotic velocity field behind the shock, the asymptotic solution can be easily obtained, by matching the downstream asymptotic velocities with the asymptotic expressions at the moving shock wave. The procedure is similar to the calculations shown in \cite{17} and will not be repeated here. For the longitudinal rotational part we get:
\begin{equation}
\tilde{v}_x^{rot}(x\gg\lambda_y,y) \cong  \left \{ \begin{array}{ll}
	Q_{rot}^l \cos{\left(Rk_xx -\phi_{rot}\right)}\cos{\left(k_yy \right)} ,\ \ \ \ \ \ \ \ \ \ \ \ \zeta_0 \leq 1 \\
				 \\
	Q_{rot}^s \cos{\left(Rk_xx\right)}\cos{\left(k_yy \right)} , \ \ \ \ \ \ \ \ \ \ \ \ \ \ \ \ \ \ \ \ \zeta_0 \geq 1  \\
	\end{array} \right.  \ \ ,
\label{38}
\end{equation}
and for the transverse component:
\begin{equation}
\tilde{v}_y^{rot}(x\gg\lambda_y,y) \cong  \left \{ \begin{array}{ll}
	R\frac{k_x}{k_y}Q_{rot}^l \sin{\left(Rk_xx -\phi_{rot}\right)}\sin{\left(k_yy \right)} ,\ \ \ \ \ \ \ \ \ \ \zeta_0 \leq 1 \\
				 \\
	R\frac{k_x}{k_y}Q_{rot}^s \sin{\left(Rk_xx\right)}\sin{\left(k_yy \right)} , \ \ \ \ \ \ \ \ \ \ \ \ \ \ \ \ \ \ \zeta_0 \geq 1  \\
	\end{array} \right.  ,
\label{39}
\end{equation}
where it is assumed that the point $x$ is far enough form the interface $x=0$ and from the shock front $[x\ll (D-U)t]$. The quantities $Q_{rot}^l$ and $Q_{rot}^s$ are:
\begin{equation}
Q_{rot}^l = \frac{\sqrt{\left(  \Omega_3  + \Omega_2 e_{lr}  \right)^2 + \left(  \Omega_2 e_{li}  \right)^2}}{1+ \left(  R \frac{k_x}{k_y}  \right)^2}
\ \ ,\ \ Q_{rot}^s = \frac{\Omega_3 + \Omega_2 e_s}{1 + \left(  R \frac{k_x}{k_y}   \right)^2} \label{40}
\end{equation}
and
\begin{equation}
\tan \phi_{rot} = \frac{\Omega_2 e_{li}}{\Omega_3 + \Omega_2 e_{lr}}
\ . \label{42}
\end{equation}
The acoustic velocity field is:
\begin{equation}
\tilde{v}_x^{ac}(x\gg\lambda_y,y,\tau\gg1) = Q_{ac}\cos{(\zeta_1 \tau - k_x^{ac} x)}\cos{(k_yy)}
\ , \label{43}
\end{equation}
\begin{equation}
\tilde{v}_y^{ac}(x\gg\lambda_y,y,\tau\gg1) = \frac{e_s}{\zeta_1}\sin{(\zeta_1 \tau - k_x^{ac} x)}\sin{(k_yy)}
\ , \label{44}
\end{equation}
where:
\begin{equation}
Q_{ac} = \frac{M_2\zeta_0-\sqrt{\zeta_0^2-1}}{\zeta_0-M_2\sqrt{\zeta_0^2-1}}e_s
\ , \label{45}
\end{equation}
only valid for the short wavelength regime $(\zeta_0>1)$.
\section{Interaction of a planar shock with a 2D random density field}
In this section, we study the interaction of an initially planar shock wave with a 2D isotropic random density field. We get averages of the downstream kinetic energy and vorticity generation. We also obtain the density amplification and the acoustic energy flux emitted by the shock. As we have seen in Section 2, the single mode profile is characterized by the vector $\vec{k}=(k_x,k_y)$. Its components are:
\begin{eqnarray}
k_x=k\cos\theta
\ , \nonumber \\
k_y=-k\sin\theta
\ , \label{47}
\end{eqnarray}
where we define $k=|\vec{k}|$, and $\theta$ is the incidence angle of the upstream perturbation wave vector with respect to the axis perpendicular to the shock front $(\hat{x})$. The range of variation is $0\leq \theta \leq \pi$.
We consider an isotropic profile and we assume that the wavenumber vector $\vec{k}$ is uniformly distributed along the unit semicircle, i.e. the probability of a particular orientation is $d\theta/\pi$. We can write the dimensionless frequency $\zeta_0$ as a function of the incident angle $\theta$ as follows:
\begin{equation}
\zeta_0=\frac{M_2 R}{\sqrt{1-M_2^2}}\frac{1}{|\tan\theta|}
\, \label{48}
\end{equation}
\subsection{Turbulent kinetic energy generation}
\begin{figure} [ht]
\begin{center}
\vspace*{-3mm}
\includegraphics[width= 0.9 \textwidth, clip]{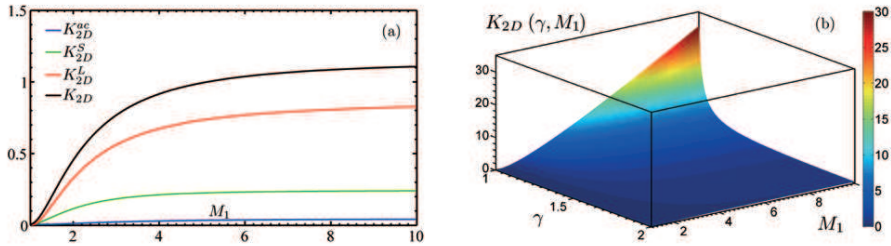}
\end{center}
\vspace*{- 6 mm}
\caption{(a) Turbulent kinetic energy generated downstream separated in its different contributions as a function of the shock Mach number for an ideal gas with $\gamma=5/3$. (b) Total kinetic energy generation as a function of $\gamma$ and $M_1$.}
\label{Figure-3}
\end{figure}

In the previous section, we have obtained the asymptotic velocity profiles downstream for a single-mode pre-shock density field. Those profiles are useful to calculate statistical averages over a full spectrum of pre-shock non-uniformities. Let us introduce a dimensionless conversion coefficient $K$ between the pre-shock density non-uniformity and the post-shock turbulent Mach number $(M_{turb})$:
\begin{equation}
M_{turb}^2=\frac{\tilde{v}^2}{c_2^2}=\left(\frac{\delta \rho_1}{\rho_1}\right)^2 \times K= K \times \epsilon_k^2
\, \label{49}
\end{equation}
where $\tilde{v}^2=|\tilde{\vec{v}}|^2=\tilde{v}_x^2+\tilde{v}_y^2$, and for the 2D problem:
\begin{equation}
K_{2D}=\frac{2}{\pi}\int_0^{\pi/2}\left
(\langle \tilde{v}_x^2\rangle+\langle \tilde{v}_y^2\rangle\right)d\theta
\, \label{50}
\end{equation}
The polar angle $\theta$ can be expressed as a function of $\zeta_0$:
\begin{equation}
d\theta=\frac{M_1\sqrt{R(M_1^2-1)}}{M_1^2R+(M_1^2-1)\zeta_0^2}d\zeta_0
\ , \label{52}
\end{equation}
and the integrals over $\theta$ can be changed into integrals over $\zeta_0$. The boundary between long and short wavelengths is given by $\zeta_0=1$ which corresponds to $\theta=\theta_{cr}$, given by \cite{17}:
\begin{equation}
\sin{\theta_{cr}}=M_1^2\sqrt{\frac{\gamma+1}{2\gamma M_1^4+(3-\gamma)M_1^2-2}}
\ . \label{53}
\end{equation}
We can decompose $K$ as the following sum: $K=K_{2D}^{l}+K_{2D}^{s}+K_{2D}^{ac}$.
where using (\ref{52}) and taking into account Eqs.(\ref{38}), (\ref{39}), (\ref{43}) and (\ref{44}), we write:
\begin{eqnarray}
K_{2D}^s&=&\frac{2}{\pi}\int_{0}^{\theta_{cr}}|\tilde{v}^{rot}|^2d\theta=\frac{2}{\pi}\int_{1}^{\infty}\left[1+\frac{(1-M_2^2)\zeta_0^2}{M_2^2}\right]|Q_{rot}^s|^2\frac{M_1\sqrt{R(M_1^2-1)}}{M_1^2R+(M_1^2-1)\zeta_0^2}d\zeta_0
\ , \nonumber \\
K_{2D}^l&=&\frac{2}{\pi}\int_{\theta_{cr}}^{\pi/2}|\tilde{v}^{rot}|^2d\theta=\frac{2}{\pi}\int_{0}^{1}\left[1+\frac{(1-M_2^2)\zeta_0^2}{M_2^2}\right]|Q_{rot}^l|^2\frac{M_1\sqrt{R(M_1^2-1)}}{M_1^2R+(M_1^2-1)\zeta_0^2}d\zeta_0
\ , \nonumber \\
K_{2D}^{ac}&=&\frac{2}{\pi}\int_{0}^{\theta_{cr}}|\tilde{v}^{ac}|^2d\theta=\frac{2}{\pi}\int_{1}^{\infty}\left(\frac{\zeta_1^2}{\zeta_1^2-1}\right)|Q_{ac}|^2\frac{M_1\sqrt{R(M_1^2-1)}}{M_1^2R+(M_1^2-1)\zeta_0^2}d\zeta_0
\ , \label{54}
\end{eqnarray}
In Fig.\ref{Figure-3}.a we plot the turbulent kinetic energy generated downstream separated in its different contributions as a function of the shock Mach number. We observe that the limiting value for the strong-shock limit is a function of $\gamma$. In Fig.\ref{Figure-3}.b we show the total kinetic energy generation as a function of $\gamma$ and $M_1$, in the ranges $1\leq \gamma \leq 2$, $1\leq M_1\leq 10$. It is observed that the total energy $K_{2D}$ grows unbounded in the strong shock limit $(M_1\rightarrow\infty)$ of a highly compressible gas $(\gamma\rightarrow 1)$.
\subsection{Acoustic energy flux}
\begin{figure} [ht]
\centering
\begin{center}
\vspace*{-3mm}
\includegraphics[width= 0.9 \textwidth, clip]{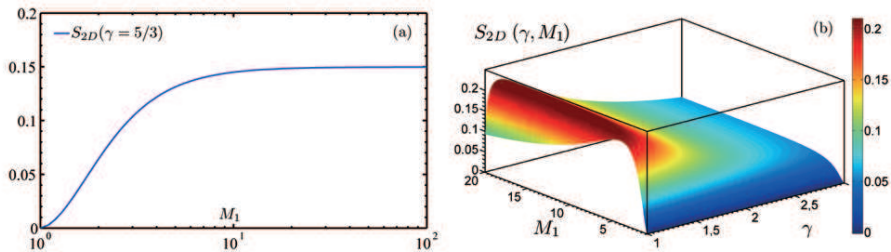}
\end{center}
\vspace*{- 6 mm}
\caption{(a) Relative acoustic energy flux $S_{2D}$ as a function of the shock strength $M_1$ for an ideal gas with $\gamma=5/3$. (b) Relative acoustic energy flux $S_{2D}$ as a function of the shock strength $M_1$ and the adiabatic index $\gamma$.}
\label{Figure-4}
\end{figure}

The emission of sound by the shock is different in the long and short wavelength regimes, as discussed in Section 2 and with more detail in \cite{16,17}. In fact, for the long-wavelength regime, the acoustic field decays exponentially following the shock and its asymptotic contribution vanishes. However, the short wavelength modes contribute with stable running fronts downstream. We study here the acoustic energy flux in two different reference systems: the compressed fluid and the shock reference frames.
The energy flux is defined by \cite{42}:
\begin{equation}
\vec{q}=c_2E\hat{k}_{ac}
\ , \label{65}
\end{equation}
where $E$ is the energy density of the sound wave:
\begin{equation}
E=\rho_2c_2^2\tilde{p}^2
\ , \label{66}
\end{equation}
in which $\tilde{p}$ is the pressure fluctuation behind the shock and can be retrieved with the aid of Eq.(\ref{24}). The unit vector $\hat{k}_{ac}$ can be expressed with the aid of Eq.(\ref{27}), and represents the direction of the sound waves emitted by the shock downstream. We have:
\begin{eqnarray}
\hat{k}_{ac} &=& \frac{k_x^{ac} \hat{x} + k_y  \hat{y}}{\sqrt{\left( k_x^{ac}\right)^2 + \left( k_y  \right)^2}} = \left( \cos \theta_{ac},  \sin \theta_{ac}\right) 	= \nonumber \\
&=&\left( \frac{M_2 \zeta_0 - \sqrt{\zeta_0^2-1}}{\zeta_0 - M_2 \sqrt{\zeta_0^2-1}}  , \frac{\sqrt{1-M_2^2}}{\zeta_0 - M_2 \sqrt{\zeta_0^2-1}}\right)
\  . \label{67}
\end{eqnarray}
The projection of Eq.(\ref{65}) along the longitudinal direction gives:
\begin{equation}
q_x=\rho_2c_2^3\tilde{p}^2\cos{\theta_{ac}}=\rho_2c_2^3e_s^2\cos{\theta_{ac}}
\  .\label{68}
\end{equation}
For simplicity, we define the dimensionless longitudinal energy flux as:
\begin{equation}
\tilde{q}_x=\frac{q_x}{\rho_2c_2^3}=e_s^2\cos{\theta_{ac}}
\ , \label{69}
\end{equation}
where $\zeta_0$ can be expressed as a function of the acoustic angle $\theta_{ac}$
\begin{equation}
\zeta_0 = \frac{1}{\sqrt{1-M_2^2}} \left(  \frac{1- M_2\cos \theta_{ac}}{\sin\theta_{ac}} \right)
\ . \label{70}
\end{equation}
In the new variable  $\theta_{ac}$, the region which corresponds to sonic waves traveling to the right is $\cos^{-1}M_2\leq\theta_{ac}\leq\pi/2$ $(1\leq\zeta_0\leq1/\sqrt{1-M_2^2})$, and the corresponding to left-facing waves is $\pi/2\leq\theta_{ac}\leq\pi$ $(1/\sqrt{1-M_2^2}\leq\zeta_0\leq\infty)$. We note that $\tilde{q}_x=0$ for $\theta_{ac}=\pi/2$ which is trivial, and also vanishes for $\zeta_0=\zeta_0^*=M_1/\sqrt{M_1^2-1}$, because $e_s(\zeta_0^*)=0$.
As has been done for the kinetic energy, we make the corresponding averages of the sonic flux over the dimensionless frequency $\zeta_0$ in the compressed fluid frame. The acoustic energy flux can be separated depending on the direction of the traveling fronts:
\begin{eqnarray}
 \langle\tilde{q}_x^{right}\rangle_{\theta} &=& \int_1^{1/\sqrt{1-M_2^2}}\tilde{q}_x(\zeta_0)\ \frac{M_1\sqrt{R(M_1^2-1)}}{M_1^2 R+ \left( M_1^2-1  \right) \zeta_0^2} d\zeta_0 \ , \nonumber \\
 \langle\tilde{q}_x^{left}\rangle_{\theta} &=& \int_{1/\sqrt{1-M_2^2}}^{\infty}\tilde{q}_x(\zeta_0)\  \frac{M_1\sqrt{R(M_1^2-1)}}{M_1^2 R+ \left( M_1^2-1  \right) \zeta_0^2} d\zeta_0
\ . \label{72}
\end{eqnarray}
It is also interesting to express the sound energy flux in the shock reference frame, because in some experiments the shock remains steady with respect to the laboratory walls, and it is the fluid upstream that moves toward the shock. In that case, the normal to the sound wave front (which coincides with $\hat{k}_{ac}$) does not coincide with the direction of propagation of the energy. The energy flux is now given by:
\begin{equation}
\vec{q}=\vec{v}_{ac}E_s
\ , \label{73}
\end{equation}
where $\vec{v}_{ac}=c_2\hat{k}_{ac}+(U-D)\hat{x}$, and $E_s$ is the energy density in the shock frame, it is written as \cite{42}:
\begin{equation}
E_s=(M_2-\cos{\theta_{ac}})E
\ , \label{74}
\end{equation}
where $E$ is given by Eq.(\ref{66}). Collecting these results we obtain:
\begin{equation}
\tilde{q}_x=\frac{q_x}{\rho_2c_2^3}=e_s^2(1-M_2\cos{\theta_{ac}})(M_2-\cos{\theta_{ac}})
\  .\label{75}
\end{equation}
In the shock reference frame, the angle between $\vec{v}_{ac}$ and the \textit{$\hat{x}$}-axis is denoted by $\theta'$. Its relation to $\theta_{ac}$ can be seen to be given by:
\begin{equation}
\cos \theta' = \frac{\cos \theta_{ac} - M_2}{\sqrt{1+M_2^2-2M_2 \cos \theta_{ac}}}
\ . \label{76}
\end{equation}
The averaged acoustic energy flux is given by:
\begin{equation}
I_{ac}=\frac{2}{\pi}\int_0^{\theta_{cr}}\tilde{q}_x d\theta
\ , \label{77}
\end{equation}
and we define the kinetic energy incident through the shock as:
\begin{equation}
I_{kin}=\rho_1\frac{D^3}{2}=\frac{1}{2}\rho_2 c_2^3 M_2^3 R^3
\ , \label{78}
\end{equation}
The dimensionless emission coefficient $S$, which connects the incident kinetic energy flux with the acoustic energy flux emitted by the shock is:
\begin{equation}
S= \frac{I_{ac}}{I_{kin}}\times \epsilon_k^2
\ , \label{79}
\end{equation}
which is not difficult to particularize for the 2D case:
\begin{equation}
S_{2D}=\frac{2}{\pi}\frac{2}{M_2^3R^2}\int_1^{\infty}\tilde{q}_x(\zeta_0)\frac{M_1\sqrt{R(M_1^2-1)}}{M_1^2R+(M_1^2-1)\zeta_0^2}d\zeta_0
\ , \label{80}
\end{equation}
In Fig.\ref{Figure-4}.a we plot the relative acoustic energy flux $S_{2D}$ as a function of the shock strength for an ideal gas with $\gamma=5/3$. In the strong shock limit $(M_1\rightarrow \infty)$, $S_{2D}$ reaches an asymptotic value dependent on $\gamma$. In Fig.\ref{Figure-4}.b we show 3D-plot for the same quantity $S_{2D}$ as a function of the shock strength $M_1$ and the adiabatic exponent $\gamma$.
\subsection{Density amplification downstream}
\begin{figure} [ht]
\centering
\begin{center}
\vspace*{-3mm}
\includegraphics[width= 0.9 \textwidth, clip]{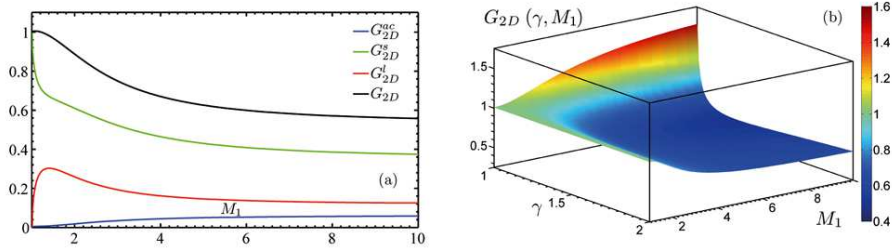}
\end{center}
\vspace*{- 6 mm}
\caption{(a) Density amplification downstream separated in its different contributions as a function of the shock Mach number for an ideal gas with $\gamma=5/3$. (b) Density amplification $G_{2D}$ as a function of the shock strength $M_1$ and the adiabatic exponent $\gamma$.}
\label{Figure-5}
\end{figure}

An important point regards the amplification of the pre-shock density perturbation field due to the shock compression. Similarly as with other perturbation quantities, the density perturbation field downstream can be decomposed as the superposition of a steady and non steady contribution:
\begin{equation}
\tilde{\rho}(k_yx,\tau)=\tilde{\rho}_{ac}(k_yx,\tau)+\tilde{\rho}_{en}(k_yx)
\ , \label{85}
\end{equation}
where the acoustic term is given by Eq.(\ref{28}):
\begin{equation}
\tilde{\rho}_{ac}(k_yx,\tau)=\tilde{p}(k_yx,\tau)
\ , \label{86}
\end{equation}
and the entropic term can be obtained with the aid of Eq.(\ref{13}) after subtracting the acoustic contribution:
\begin{equation}
\tilde{\rho}_{en}(k_yx)=\left(\frac{1}{M_1^2 M_2^2}-1\right)\tilde{p}_s\left(\frac{k_y x \sqrt{1-M_2^2}}{M_2}\right)+\frac{\delta\rho_1(k_yx)}{\rho_1}
\ . \label{87}
\end{equation}
We define the average of the ratio between the downstream over upstream asymptotic density field as follows:
\begin{equation}
\frac{\langle\delta\tilde{\rho}_2^2\rangle}{\langle\delta\tilde{\rho}_1^2\rangle}=\frac{1}{R^2}\frac{\langle\delta\rho_2^2\rangle}{\langle\delta\rho_1^2\rangle}= G
\ . \label{88}
\end{equation}
As for the dimensionless function $G$, we particularize for the 2D case, and explicitly calculate the long/short wavelength and the acoustic contributions:
\begin{eqnarray}
G_{2D}^l &=& \frac{2}{\pi}\int_{\theta_{cr}}^{\pi/2}\left[\frac{(1-M_1^2M_2^2)^2}{M_1^4M_2^4}(e_{lr}^2+e_{li}^2)+\frac{2(1-M_1^2M_2^2)}{M_1^2M_2^2}e_{lr}+1\right]d\theta\ , \nonumber \\
G_{2D}^s &=& \frac{2}{\pi}\int_{0}^{\theta_{cr}}\left[\frac{(1-M_1^2M_2^2)^2}{M_1^4M_2^4}e_{s}^2+\frac{2(1-M_1^2M_2^2)}{M_1^2M_2^2}e_{s}+1\right]d\theta\  , \nonumber \\
G_{2D}^{ac} &=& \frac{2}{\pi}\int_{0}^{\theta_{cr}}e_s^2d\theta\
\ . \label{89}
\end{eqnarray}
We plot these quantities for a gas with $\gamma=5/3$ in Fig.\ref{Figure-5}.a, and we observe that $G_{2D}^s>G_{2D}^l$, in contrast with the behavior in Fig.\ref{Figure-3}, where $K_{2D}^s<K_{2D}^l$. The total density amplification is shown in a 3D-plot in Fig.\ref{Figure-5}.b as a function of $\gamma$ and $M_1$.
\subsection{Vorticity generation downstream}
\begin{figure} [ht]
\centering
\begin{center}
\vspace*{-3mm}
\includegraphics[width= 0.9 \textwidth, clip]{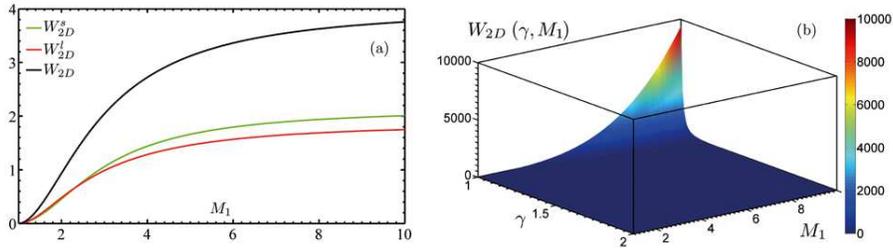}
\end{center}
\vspace*{- 6 mm}
\caption{(a) Vorticity generated downstream separated in its different contributions as a function of the shock Mach number for an ideal gas with $\gamma=5/3$. (b) Vorticity generation $W_{2D}$ as a function of the shock strength $M_1$ and the adiabatic index $\gamma$.}
\label{Figure-6}
\end{figure}

We focus now on the vorticity generated by the shock oscillations. This vorticity is conserved for each fluid element after shock compression in the absence of viscosity. Therefore, in the compressed fluid frame the vorticity is steady. Using (\ref{31}), we have for any single-mode perturbation:
\begin{equation}
\tilde{\omega}_z=\frac{\omega_z}{k_yc_2}=\frac{\omega_z}{|\vec{k}|c_2\sin{\theta}}=\Omega_2 \tilde{p}_s+\Omega_3\frac{\delta\rho_1}{\rho_1}
\ , \label{100}
\end{equation}
where $\Omega_2$ and $\Omega_3$ have been defined in Eqs.(\ref{32}) and (\ref{33}). In order to obtain the corresponding averages, it is convenient to use $k=|\vec{k}|$ for the dimensionless vorticity $\tilde{\omega}_z$. With this idea, we define this new vorticity as: $\tilde{\omega}_z=\omega_z/(k\  c_2)$. Thus, this quantity is used to get the dimensionless factor $W$, which quantifies the averaged vorticity generated by the corrugated shock:
\begin{equation}
\langle\tilde{\omega}_z^2\rangle=W \times \epsilon_k^2
\ , \label{101}
\end{equation}
Again, we particularize the averaged quantity $W$ for the 2D case, and it can be also separated in its different wavelength regimes:
\begin{eqnarray}
W_{2D}^l &=& \frac{2}{\pi}\int_{\theta_{cr}}^{\pi/2}\left[\Omega_2^2(e_{lr}^2+e_{li}^2)+2\Omega_2\Omega_3e_{lr}+\Omega_3^2\right]\sin^2{\theta}d\theta\ , \nonumber \\
W_{2D}^s &=& \frac{2}{\pi}\int_{0}^{\theta_{cr}}\left[\Omega_2^2e_{s}^2+2\Omega_2\Omega_3e_{s}+\Omega_3^2\right]\sin^2{\theta}d\theta\
\ , \label{102}
\end{eqnarray}
In Fig.\ref{Figure-6}.a, we plot $W_{2D}^{l}$, $W_{2D}^{s}$ and $W_{2D}=W_{2D}^{l}+W_{2D}^{s}$. In Fig.\ref{Figure-6}.b, we show the total vorticity generated behind the shock, for different values of $\gamma$ and $M_1$. We see that $W_{2D}$ is divergent for $M_1\gg1$ and $\gamma-1\ll1$. This is because of the size reduction of the eddies in that limit. This divergent behavior would be smoothed in a real gas, because, viscosity effects would become important for strong shocks traveling in highly compressible gases.

\section{Summary}
An analytical linear theory for the interaction of a planar shock wave with a 2D random density field has been presented. The entropy spectrum is assumed to be isotropic. First, the simpler problem of the interaction with a 2D single-mode perturbation has been developed, in which, the acoustic waves and the vorticity/entropy perturbations generated downstream have been obtained.Thanks to the isotropy of the upstream perturbations, the important statistical averages downstream can be easily calculated. The turbulent kinetic energy and vorticity generation behind the shock are obtained as a function of the shock strength $M_1$, and the gas compressibility $\gamma$. The acoustic energy flux (noise) emitted by the shock front is studied in the compressed fluid and shock reference frames. The amplification of the density perturbation across the shock is also discussed. The perturbation fields generated downstream consist of vorticity/entropy structures which induce mass and momentum redistribution and its associated mixing process. The results shown here can be used to deal with other types of interactions (random acoustic field or isotropic vorticity perturbations). 3D isotropic spectra can also be studied with the tools shown here, by changing the probability density function for the 3D case as done in \cite{17}. The model presented here can also be extended to account different boundary conditions, such as the presence of a piston, a free surface or an ablation front driving the shock. Besides, it might also be applied to study the effect if a second shock traveling into the compressed spectrum generated by the first one.


\section{References}
\begin{thebibliography}{10}

\bibitem{1}R. Richtmyer, Commun. Pure Appl. Math. {\bf 13}, 297 (1960).

\bibitem{2}E. E. Meshkov, Fluid Dyn. {\bf 4}, 101 (1969).

\bibitem{3}G. Fraley, Phys. Fluids {\bf 29}, 376 (1986).

\bibitem{4}P. M. Zaidel, J. Appl. Math. Mech. {\bf 24}, 316 (1960).

\bibitem{5}H. S. Ribner, N. A. C. A. Rep. 1164 (1954).

\bibitem{6}H. S. Ribner, N. A. C. A. Rep. 3255 (1954).

\bibitem{7}H. S. Ribner, AIAA J. {\bf 25}, 436 (1987).

\bibitem{8}H. S. Ribner, AIAA J. {\bf 36}, 494 (1998).

\bibitem{9}H. S. Ribner, J. Fluid Mech. {\bf 35}, 299 (1969).

\bibitem{10}F. K. Moore, N. A. C. A. Rep. 2879 (1953).

\bibitem{11}J. L. Kerrebrock, Ph. D. Thesis, California Institute of Technology (1956).

\bibitem{12}D. Rotman, Phys. Fluids {\bf 3}, 1792 (1991).

\bibitem{13}K. Mahesh, S. Lee, S. K. Lele, and P. Moin, J. Fluid Mech. {\bf 300}, 383 (1995).

\bibitem{14}K. Mahesh, S. K. Lele, and P. Moin, J. Fluid Mech. {\bf 334}, 353 (1997).

\bibitem{15}K. Mahesh, Ph. D. Thesis, California Institute of Technology (1996).

\bibitem{16}A. L. Velikovich, J. G. Wouchuk, C. Huete Ruiz de Lira, N. Metzler, S. Zalesak, and A. J. Schmitt, Phys. Plasmas {\bf 14}, 072706 (2007).

\bibitem{17}J. G. Wouchuk, C. Huete Ruiz de Lira, A. L. Velikovich, Phys. Rev. E {\bf 79}, 066315 (2009).

\bibitem{18}N. K.-R. Kevlahan, J. Fluid Mech. {\bf 327}, 161 (1996).

\bibitem{19}N. K.-R. Kevlahan, J. Fluid Mech. {\bf 341}, 371 (1997).

\bibitem{20}S. Lee, S. K. Lele, and P. Moin, J. Fluid Mech. {\bf 340}, 225 (1997).

\bibitem{21}S. Lee, S. K. Lele, and P. Moin, J. Fluid Mech. {\bf 251}, 533 (1993).

\bibitem{22}G. Dimonte, and R. Tipton, Phys. Fluids {\bf 18}, 85101 (2006).

\bibitem{23}S. Barre, D. Alem, and J. P. Bonnet, AIAA J. {\bf 34}, 968 (1996).

\bibitem{24}S. Barre, D. Alem, and J. P. Bonnet, AIAA J. {\bf 36}, 495 (1998).

\bibitem{24b}J. Keller and W. Merzkirch, Exp. Fluids {\bf 8}, 241 (1990).

\bibitem{25}A. L. Velikovich, Phys. Fluids {\bf 8}, 1666 (1996).

\bibitem{26}J. G. Wouchuk, Phys. Rev. E {\bf 63}, 056303 (2001).

\bibitem{27}J. G. Wouchuk, and J. Lopez Cavada, Phys. Rev. E {\bf 70}, 046303 (2004).

\bibitem{28}A. D. Kotelnikov and D. C. Montgomery, Phys. Fluids {\bf 10}, 2037 (1998).

\bibitem{29}N. Metzler, A. L. Velikovich, and J. H. Gardner, Phys. Plasmas. {\bf 9}, 5050 (2002)

\bibitem{30}F. Philippe, B. Canuad, X. Fortin, F. Garaude, and H.Jourdren, Laser Part. Beams. {\bf 22}, 171 (2004)

\bibitem{31}G. Hazak, A. L. Velikovich, J. H. Gardner and J. P. Dahlburg, Phys. Plasmas {\bf 5}, 4357 (1998).

\bibitem{32}J. B. Collins, A. Poludnenko, A.Cunningham, and A. Frank, Phys. Plasmas. {\bf 12}, 062705 (2005)

\bibitem{33}J. D. Moody, B. J. MacGowan, S. H. Glenzer, R. K. Kirkwood, W. L. Kruer, D. S. Montgomery, A. J. Schmitt, E. A. Williams, and G. F. Stone, Phys. Plasmas. {\bf 7}, 2114 (2000)

\bibitem{34}M. Desselberger, M. W. Jones, J. Edwards, M. Dunne and O. Willi, Phys. Rev. Lett. {\bf 74}, 2961 (1995)

\bibitem{35}R. J. Manson, R. A. Kopp, H. X. Vu, D. C. Wilson, S. R. Goldman, R. G. Watt,  M. Dune, and O. Willi, Phys. Plasmas. {\bf 5}, 211 (1998)

\bibitem{36}B. A. Remington, R. P. Drake, H. Takabe, and D. Arnett, Phys. Plasmas. {\bf 7}, 1641 (2000)

\bibitem{37}A. Y. Poludnenko, A. Frank and E. G. Blackman, Astrophys. J. {\bf 576}, 832 (2002).

\bibitem{37b}N. K.-R. Kevlahan, and R. E. Pudritz, Astrophys. J. {\bf 702}, 39 (2009).

\bibitem{38}W. R. LePage, {\it Complex Variables and the Laplace Transform for Engineers} (Dover, New York, 1980).

\bibitem{39}B. Davies, {\it Integral transforms and their applications} (Springer, New York, 1984).

\bibitem{40}I. S. Gradshteyn, and I. M. Ryzhik, {\it Table of Integrals, Series, and Products}, $5^{th}$ Edition (Academic Press, San Diego, 1994).

\bibitem{42}L. D. Landau, and E. M. Lifshitz, {\it Fluid Mechanics} (Pergamon Press, New York, 1987).


\end{thebibliography}
\end{document}